\documentclass[aps,prl,showpacs,twocolumn]{revtex4}

\usepackage{amsmath,amssymb,graphicx}

\newcommand{\be}{\begin{equation}}
\newcommand{\ee}{\end{equation}}
\newcommand{\bea}{\begin{eqnarray}}
\newcommand{\eea}{\end{eqnarray}}
\newcommand{\non}{\nonumber}

\begin{document}
\title{Particlelike distributions 
of the Higgs field nonminimally coupled to gravity}

\author{Andr\'e F\"uzfa}
\email{andre.fuzfa@unamur.be}
\author{Massimiliano Rinaldi}
\email{massimiliano.rinaldi@unamur.be}
\author{Sandrine Schl\"ogel}
\email{sandrine.schlogel@unamur.be}
\affiliation{Namur Center for Complex Systems (naXys), University of Namur, \\
Rue de Bruxelles 61, B-5000 Namur, Belgium}

\date{\today}

\begin{abstract}
\noindent 
When the Higgs field is nonminimally coupled to gravity, there exists
a family of spherically symmetric particlelike solutions to the field equations. 
These monopoles are the only globally regular and asymptotically flat distributions with finite energy of the Higgs field around compact objects. Moreover, spontaneous scalarization is strongly amplified for specific values of their mass and compactness.
 \end{abstract}

\pacs{04.50.Kd,04.25.D-,04.40.-b}

\maketitle

\noindent General relativity (GR) and the standard model  (SM)  of particle physics provide for two rather different concepts of mass. While the first is rooted in the geometrical properties of space-time, the second relates the mass of elementary particles to their interactions with the fundamental Higgs scalar field. Generally speaking, scalar fields are also important for gravitational physics, as they might explain cosmic acceleration both in the early Universe (inflation) and in the late one (dark energy). In particle physics, the Higgs field offers a successful description of electroweak symmetry breaking, but it still  relies upon a  classical dynamical mechanism to settle to its vacuum expectation value (VEV). Several attempts were made to combine the Higgs mechanism with gravitation, mainly by considering the Higgs field as a partner of the metric in mediating gravitational interactions. This is the key idea of the scalar-tensor theories of gravity, developed from the pioneering work of Brans and Dicke \cite{brans}, where the scalar field induces variations of the gravitational coupling, leading to a violation of the Einstein equivalence principle. 

Considering the Higgs field as a nonminimally coupled partner of gravity has recently lead to promising results in inflationary cosmology \cite{shapo1}. In fact, the nonminimal coupling flattens the effective inflationary potential for large field values so that there is a sufficiently long slow-rolling phase that terminates when the Higgs field starts oscillating around its VEV. This model simply identifies the inflaton with the Higgs field, although concerns about the loss of unitarity lead one to consider additional scalar fields and unimodular gravity \cite{unimodular,unitarity}. In addition, Higgs inflation is among the most favored inflationary models according to the Planck results \cite{planck}. 

The nonminimal coupling constant $\xi$ is related to the slow-roll parameters so it is constrained by observations, which indicates that $\xi$ is very large, of the order of $10^{4}$ \cite{shapo1}. This naturally raises concerns about static configurations: how a such strongly coupled Higgs field reacts in the presence of gravitationally bound matter? What does the vacuum look like in the vicinity of a compact object? Since the works of Damour and Esposito-Far\`ese \cite{spontscal}, we know that a nonminimally coupled scalar field can give rise to spontaneous scalarization in compact objects. Along these lines, in this Letter we show that all spherically symmetric distributions of matter carry a classical Higgs charge, whose magnitude depends on their mass, their compactness, and the strength of  $\xi$.  To do so, we study the spherically symmetric and static solutions of the equations of motion derived from the Lagrangian, written in Jordan frame  \footnote{We choose the mostly plus signature, so the potential and the kinetic term have the same sign, in opposition to the choice of ref.\ \cite{shapo1} and in agreement with ref.\ \cite{unimodular}.}
\bea\non
{\cal L}&=&\sqrt{-g}\Bigg[\left(1+\xi v^{2}h^{2}\right){M^{2}R\over 16\pi}-{M^{2}v^{2}\over 2}(\partial h)^{2}\\
&-&{\ell M^{4}v^{4}\over 4}(h^{2}-1)^{2}\Bigg]+{\cal L}_{\rm m}(g_{\mu\nu};\Psi),\label{lagra}
\eea
where $R$ is the Ricci scalar, $M$ is the Planck mass, $\ell\sim 0.1$ is the SM coupling of the Higgs potential, $h$ is the dimensionless Higgs field in unitary gauge, $v=246\, {\rm GeV}/M=2.01\times 10^{-17}$, and $\xi$ is the nonminimal coupling. We assume that the matter fields $\Psi$, coupled to the metric $g_{\mu\nu}$, form a perfect fluid with stress tensor $T^{\mu}_{(m)\nu}=$diag$(-\rho,p,p,p)$, obtained by variation of the matter Lagrangian ${\cal L}_{\rm m}$. 

The Klein-Gordon equation easily follows and reads
\be
\Box h=\left[-\frac{\xi R}{8\pi}+\ell M^2 v^2(h^2-1)\right]h\ .
\label{kg}
\ee 
The function $h(r)=0$ is always a solution, even  with nonminimal coupling and in the presence of matter, i.e. when  $R\neq 0$. However, this solution has infinite energy since the potential becomes $V(h=0)=(1/4)\ell M^4 v^4\ne 0$ so the metric describes an asymptotically de Sitter space.
If we set $\xi=0$, there is only one asymptotically flat solution
of finite energy, namely $h=1$ everywhere. With no matter, this holds also for $\xi\neq 0$ because of a well-known no-hair theorem \cite{soti}. The interesting case is the nonminimally coupled one in the presence of matter as $h=1$ cannot be a global solution. In fact, in this case there exists a family of nontrivial solutions with asymptotically flat metric and finite energy. They describe compact objects that are charged under the Higgs field, which tends to its VEV at spatial infinity.  As they behave as isolated SM scalar charges, we dub these particlelike distributions \textit{Higgs monopoles}.

To demonstrate our claim, we  integrate numerically Eq.\ \eqref{kg} together with the $(t,t)$ and the $(\theta,\theta)$ components of Einstein equations, written in Schwarzschild gauge. The $(r,r)$ component is used for a consistency check. For the sake of simplicity, we consider a top-hat distribution of matter
($\rho_m=\rho_0$ for $r\le \mathcal{R}$, $\rho_m=0$ for $r> \mathcal{R}$, $\mathcal{R}$ being the radius of the compact object). Boundary conditions are imposed by regularity of the metric and of the scalar field at the center of the body, together with the condition  $p(r=\mathcal{R})=0$.  We use a shooting method to integrate along the radial direction for different values of $h(r=0)=h_c$, which is the only free parameter left, together with the mass $m$ and the compactness $s$ of the body, defined as  $s=r_s/\mathcal{R}$, $r_s=2Gm/c^2$ being the Schwarzschild radius.
\begin{figure}[ht!]
\includegraphics[scale=0.25]{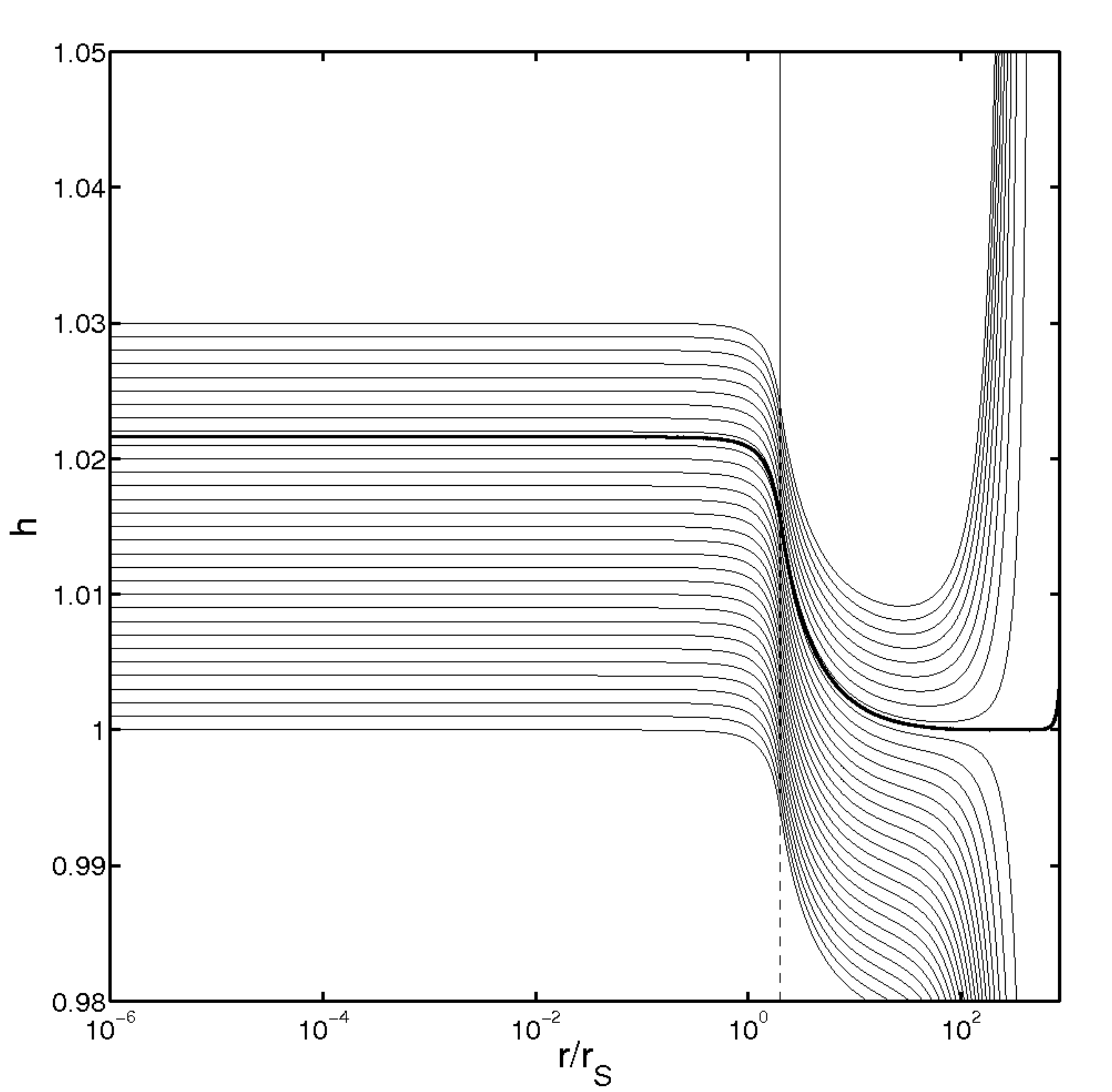}
\caption{Plot of $h(r/r_{s})$  for several values of $h_c$. The vertical line marks the radius of the body and we set $\xi=1$, $m=10^7$ kg, and $s=0.5$.
}\label{hcvsr}
\end{figure}
As we see from Fig.\ (\ref{hcvsr}), if  $h_c$ is too small, the field is attracted downwards and we checked that it converges to the infinite energy configuration $h=0$.
In the opposite case, if $h_c$ is too large, the Higgs field eventually diverges, leading again 
to infinite energy configurations. However, there exists a value of $h_{c}$ such that the Higgs field converges to its VEV at spatial infinity (thicker black line in Fig. 1) and such that the metric is asymptotically flat. This is our first important result: by requiring a finite value for the energy and global regularity, we find a configuration, the Higgs monopole, labeled by $h_{c}$ and characterized by its mass, compactness, and coupling strength $\xi$.  In Fig.\ \ref{hvsr} we plot four Higgs monopoles, found with the shooting method, with the parameters specified in Table (\ref{table}). We note that their profiles vary significantly, with rapid oscillations near the inner surface of the compact object only for certain values of $\xi$.
As the Higgs field does not vanish outside the body,  monopoles can interact with the surrounding matter both through gravitation and the SM coupling. We find that, for low compactness,  the external Higgs field can be faithfully modeled  by the Yukawa function $h(r)\approx v[(r_sQ/r)\exp(-r/(r_s L))+1]  $\footnote{In usual tensor-scalar
theories, $V=0$ so that the field decreases as $\alpha r_S/r$ with a scalar charge 
$\alpha$ proportional to $v^2Q$. In this case, one can directly relate $\alpha$ to the PPN parameters $\gamma$ and $\beta$ or  to the effective Newton constant \cite{damour}. In our case, as $V\neq 0$, the scalar field is massive and  it decreases with a Yukawa profile, which almost completely screens the 
scalar charge over a distance of few Schwarzschild radii.}. The orders of magnitude of $Q$ and $L$ for the Higgs monopoles in Fig.\ (\ref{hvsr}) are listed in table (\ref{table}).
\begin{figure}[ht!]
\includegraphics[scale=0.27,clip=true,trim= 170 5 0 40]{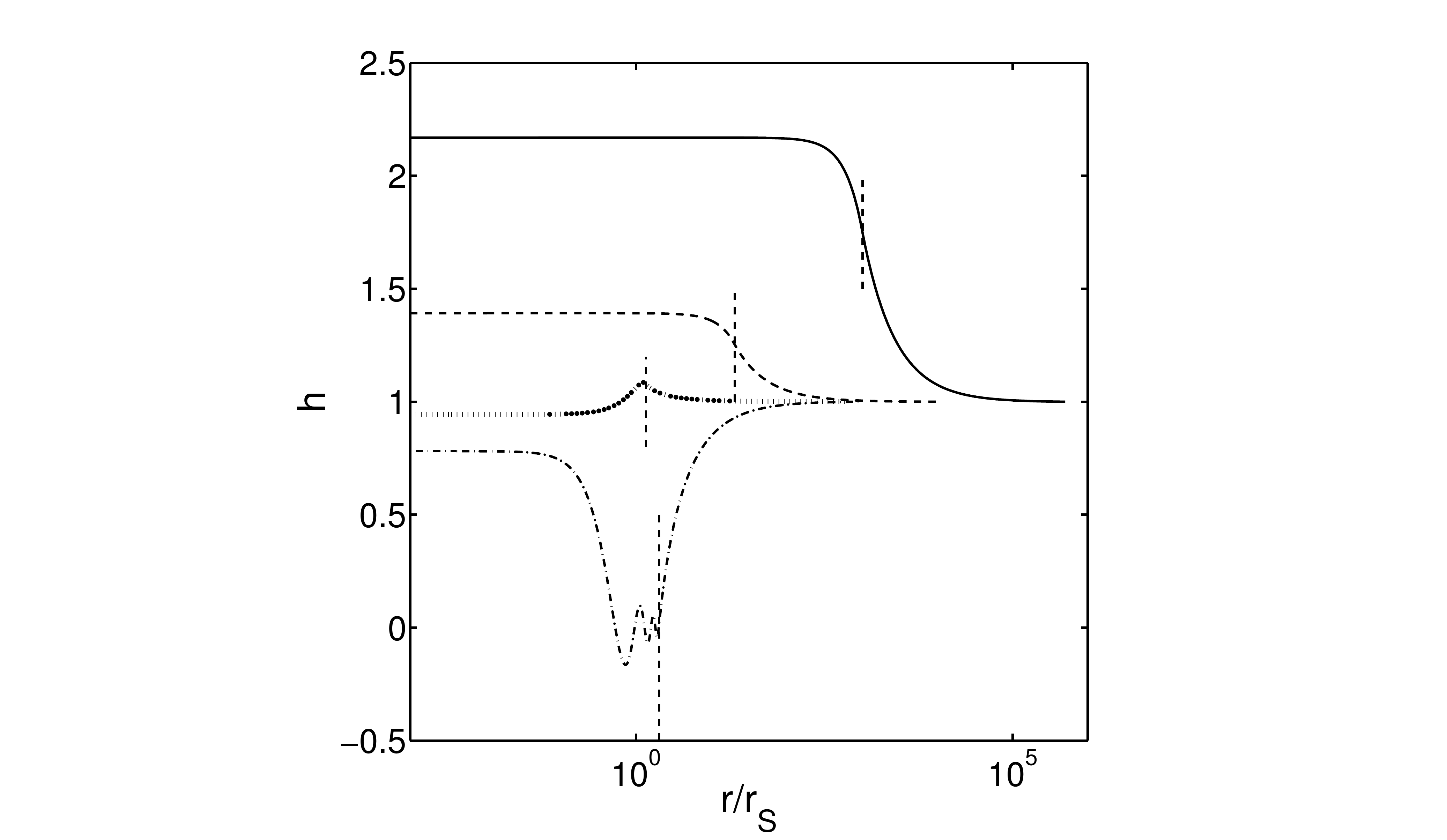}
\caption{Profile of the Higgs monopoles listed in Table (\ref{table}).
Vertical dashed lines mark the radius of the body for each monopole.
}\label{hvsr}
\end{figure}
\begin{table}[h]
\begin{tabular}{|c|c|c|c|c|c|c|}
\hline
Style & $\xi$ & m &  s &  Q & $L $ \\
\hline
solid & $10^4$ & 10\,\,TeV & $10^{-3}$ &  763 & $3.8\times 10^5$\\
\hline
dashed & $10^2$ & $2.2\times 10^2$kg & $0.05$  & 5.2 & $7.7\times 10^2$\\
\hline
dot-dashed & $10^4$ & $2.2\times 10^4$kg & $0.5$  & -1.97 & $ 5.01$\\
\hline
dotted & $10$ & 100 GeV & $0.75$ &  0.1 & 1.34\\
\hline
\end{tabular}
\caption{Properties of the Higgs monopoles plotted in Fig.\ (\ref{hvsr}). }\label{table}
\end{table}

Each monopole is also characterized by its gravitational binding energy defined by $E_{bin}=E_{m}-E_{ADM}$, where
$E_m=\int T_{0,(m)}^0\sqrt{^{(3)}g}d^3x$ is the baryonic energy 
($^{(3)}g$ is the determinant of the spatial metric) and 
$E_{ADM}= \int T_{0,(m+h)}^0 \sqrt{^{(3)}g}d^3x$ is the ADM mass
(where $T_{0,(m+h)}^0$ is the total energy density of matter plus Higgs field and the integral is evaluated at spatial infinity). When $V=0$, there are always two solutions that reduce to GR at infinity: the trivial one, with vanishing scalar field everywhere, and a family of non-trivial ones, where the compact object carries a scalar charge with asymptotically vanishing effects \cite{spontscal}, as the scalar field becomes constant at spatial infinity. The difference between the binding energies of the two configurations tells which one is the most energetically favored. However, in our case, the nonzero potential discards the trivial solution so we cannot compare it with the non-trivial one. In other words, the Higgs monopole has finite energy and it is a unique solution that cannot smoothly decay to the GR one.

\begin{figure}[ht!]
\includegraphics[scale=0.25,trim= 250 60 0 40]{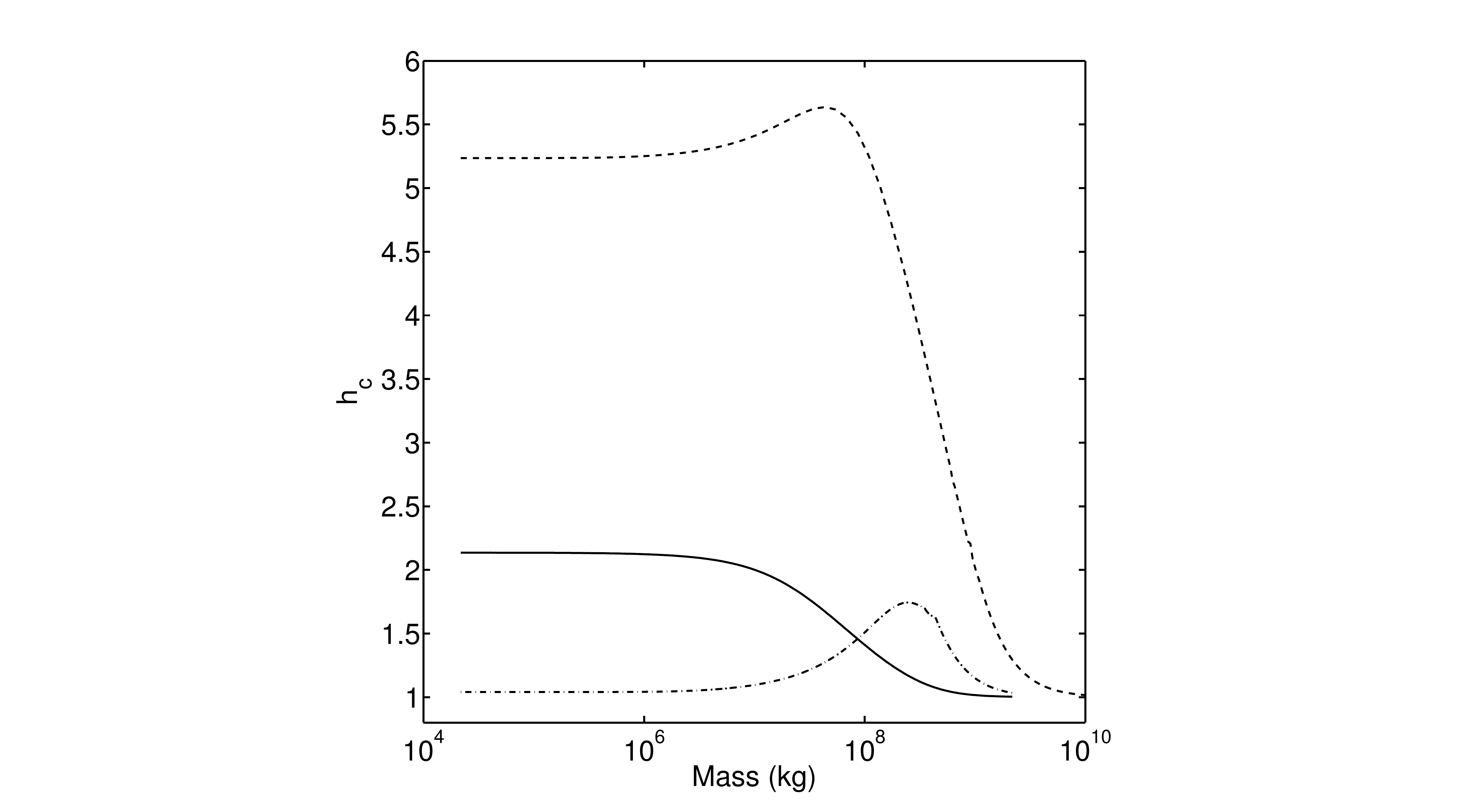}
\caption{Plot of $h_c$ as a function of $m$ with $s=0.1$ for increasing coupling: $\xi=10^2$ (solid line), $\xi=10^3$ (dot-dashed line) and  $\xi=10^4$ (dashed line).
}\label{hcvsm}
\end{figure}

Let us now look at the properties of the Higgs monopoles. In Fig.\ (\ref{hcvsm}), we show how $h_{c}$ depends on the mass for fixed compactness $s=0.1$ and for various $\xi$, including the typical value of Higgs inflation $\xi=10^4$.
We note that $h_{c}$ always exhibits a plateau at small $m$ and tends to $1$ at large $m$ values. For sufficiently large $\xi$,
$h_c$ shows a maximum before tending to 1 for large $m$ values.
This behavior can be understood if we write Eq.\ \eqref{kg} as $\square h = -{dV_{\rm eff}/ dh}$ where the effective potential $V_{\rm eff}(h)$ changes sharply at $r={\mathcal R}$. Inside 
the matter region, we find that $R$ is nearly constant and well approximated
by $R\approx R(r=\mathcal{R})=3 s^3/r_s^2$, so
\be
V_{\rm eff}^{\rm in}\simeq \frac{3\xi s^{3}  h^{2}}{16\pi r_{s}^{2}}-\frac{\ell}{4}M^2 v^2(h^2-1)^2.
\label{Veff}
\ee
The equilibrium points of $V_{\rm eff}(h)$ are 
\be
h_{\rm eq}^{\rm in}=0,\,\pm\sqrt{1+\frac{3s^3\xi}{8\pi r_s^2\ell M^2 v^2}}\cdot
\label{heq}
\ee
For $r>{\mathcal R}$, we have that $R\approx 0$ and the Higgs field is essentially driven by the quartic potential with extrema at $h_{\rm eq}^{\rm out}=0,\pm 1$. 
\begin{figure}[ht!]
\includegraphics[scale=0.3,trim= 200 180 200 150]{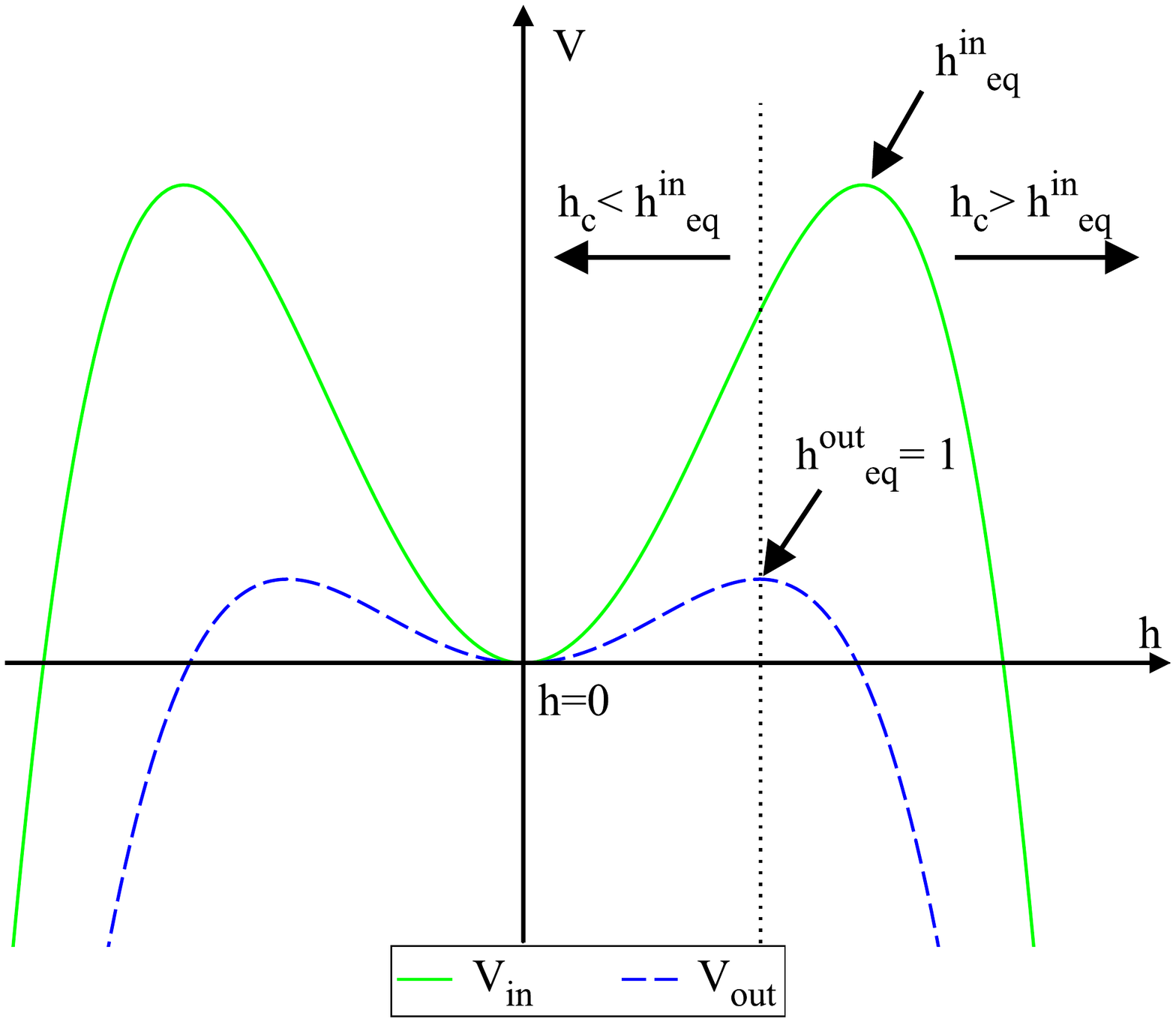}
\caption{
Qualitative representation of the effective potentials for the Higgs field inside (green solid curve) and outside 
(dashed blue curve) the body. The arrows show the directions along which the field
is initially driven.}
\label{plotVeff}
\end{figure}
In Fig.\ (\ref{plotVeff}) we show a qualitative plot for the two cases:
the field rolls down the effective potential
from some initial value $h_c$ with zero speed, as required by standard regularity conditions. If $\xi=0$, the potential does not change across the surface of the body (dashed blue
curve). Therefore, if the initial condition is $|h_c|>1$ the field will diverge at spatial infinity while, for $|h_c|<1$, it will be attracted towards $h=0$, where it eventually settles. This prevents the existence of Higgs monopoles in GR, leaving the trivial solutions $h=0$ (of infinite energy) and $h=\pm 1$ as the only solutions. However, when $\xi\ne 0$
there exist particlelike solutions that interpolate between some
value $h_c$ at the center and $h=1$ at spatial infinity. This holds only for $0<|h_c|<|h_{\rm eq}^{\rm in}|$, i.e. when the driving force pushes $h$ towards $h=1$, otherwise the field rolls down towards increasing values of $h$ and eventually diverges.  This is another important result: the Higgs monopoles exist only because of the nonminimal coupling that changes the potential across the surface of the body.

From Eq.\ \eqref{heq}, we see that  $h_c$ is bounded from above. For example,  if $\xi=10^4$ we find that
$h_c<1.01$ for a mass $m>3\times 10^{10}$ kg with $s=0.2$ and $h_c<1.01$  for 
a compactness of $s<10^{-5}$ and a mass of $10^4$ kg. We also find that the variation of the Higgs field cannot be higher than $10^{-41}v$ inside objects such as neutron stars ($s\simeq0.2$, $m\simeq 10^{30}$ kg). 

With these results, we can estimate the deviations from GR in astrophysics. We note that $|h(r)|\le |h_c|\le |h_{\rm eq}^{\rm in}|$ $\forall\,\, r\ge 0$ and that,
from Eq.\ \eqref{heq},  $h_{\rm eq}^{\rm in}-1\ll 10^{-2}$ for an object like the Sun
($m\approx 10^{30}\rm kg$ and $s\approx 10^{-6}$) provided $\xi<10^{58}$. Therefore, for astrophysical objects and reasonable values of $\xi$, we can consider $h_c\approx 1$. Following Ref.\ \cite{damour}, we then find that the
PPN parameters satisfy $\gamma-1 \ll \xi^2 v^2/\pi\approx 10^{-26}$ and
$\beta-1\ll \xi^3 v^2/(2\pi^2)\approx 10^{-23}$ for $\xi=10^4$. These are upper bounds  since the Higgs field decays as a Yukawa function outside 
the matter distribution at a much faster rate than  $1/r$ (typical in the case of a vanishing scalar potential, \cite{damour}). Therefore, we can conclude that deviations from GR around astrophysical objects like the Sun are outstandingly small, provided the coupling parameter $\xi$ is not extremely large ($\xi<10^{58}$ for $s\approx 10^{-6}$).

For small masses and fixed compactness, the first term in Eq.\ \eqref{Veff} dominates over the Higgs potential for $r<{\mathcal R}$. Then, the dynamics is similar to the one studied in \cite{spontscal}, where the field inside the body is almost constant and this explains the plateau in Fig.\  (\ref{hcvsm}). However, outside the body $R\approx 0$ and the Higgs field decreases faster than that in \cite{spontscal} because of the quartic potential. 

Let us now discuss the maxima in $h_c$ appearing in Fig.\ (\ref{hcvsm}). When $\xi$ is sufficiently large, the field undergoes damped oscillations near the inner surface of the body (see the dot-dashed curve
in Fig.\ (\ref{hvsr})). Just outside the body, the field  is  a monotonically decreasing function and  the convergence to the VEV at infinity  occurs only for precise values of the field amplitude and gradient at the boundary. In turn, this means that the amplification is possible only
 for specific values of mass, coupling, and radius of the matter distribution boundary. For a monopole of given mass, this mechanism can be seen as a classical resonance between the size of the body and
the central value of the Higgs field $h_{c}$. 
 \begin{figure}[ht!]
\includegraphics[scale=0.3,clip=true,trim=15 8 0 20]{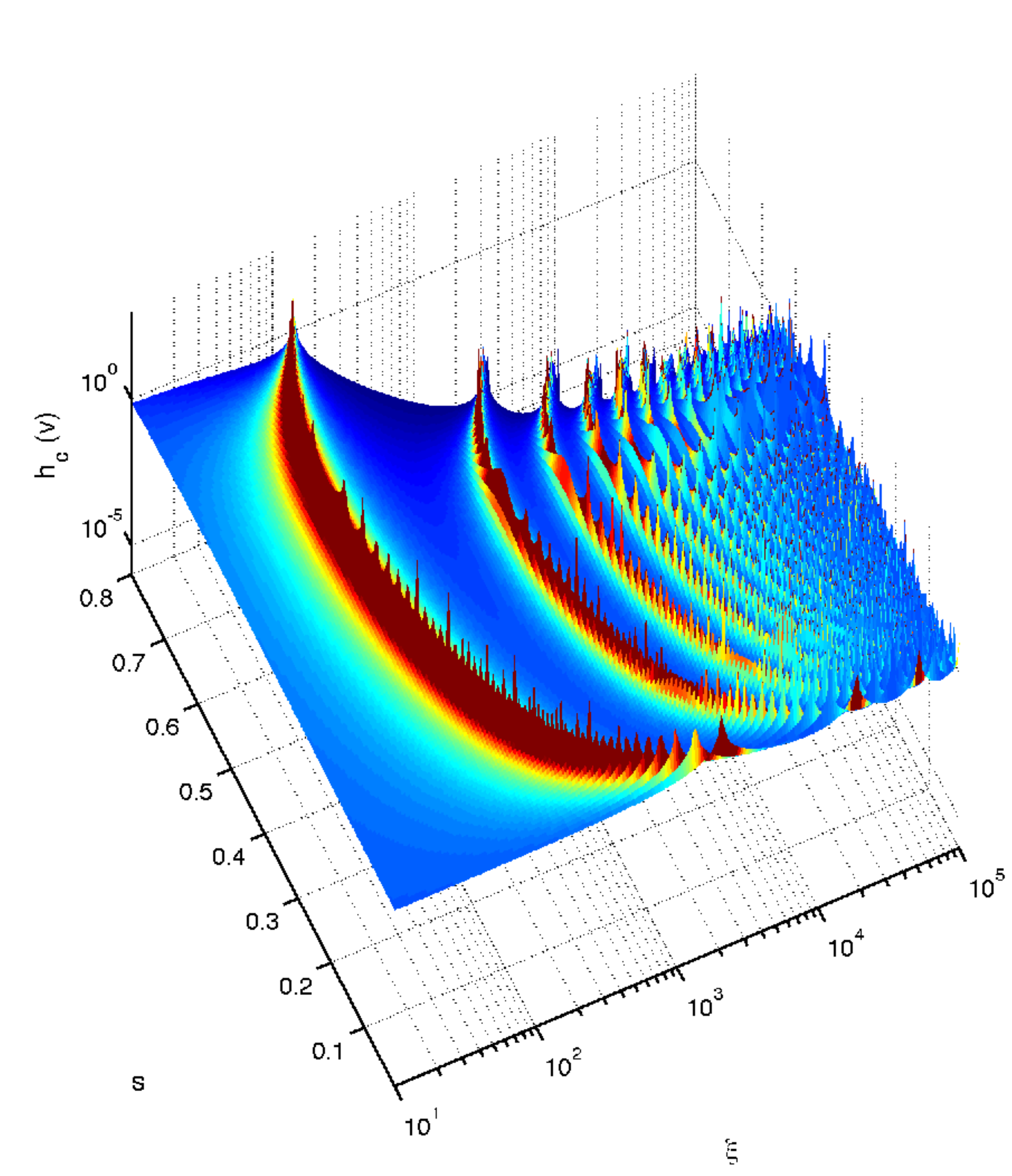}
\caption{
Plot of $h_c$ as a function of $s$ and $\xi$  for $m=10$ TeV.
}\label{plot3d}
\end{figure}

In Fig.\ (\ref{plot3d}), we plot
 $h_c$ as a function of the compactness for $\xi$ varying between $1$ and $10^{5}$.
The peaks in $h_c$ appear for  $\xi \gtrsim 100$  and are located
along sharp curves in the $(s,\xi)-$plane \footnote{In previous papers on spontaneous scalarization, these resonances were not found as  the region of large coupling was ignored.}. As $\xi$
increases, these curves get more and more crowded with resonances, and their distribution becomes chaoticlike for $\xi \gtrsim 10^5$. 
Although every point on the $(s,\xi)-$plane represents a Higgs monopole with a given mass, the scalar charge at its boundary strongly depends on  whether it is located on or not on a resonance. In fact, these resonant monopoles represent the most strongly interacting ones with matter.

In conclusion, we believe that our results can shed some light on the interplay between particle masses and gravitation. We stress once more that these solutions are impossible in GR as they exist only because of the violation of the equivalence principle. In addition, whereas other particlelike solutions  exist only in the context of exact and unbroken gauge symmetry, as in the Einstein-Yang-Mills system \cite{bartnik}, our solution is compatible with spontaneously broken gauge symmetry at the price of a nonminimal coupling to gravity.  In Higgs inflation, these monopoles could form and if they are not  washed out by the exponential expansion, they could constitute a candidate for dark matter, with a mass range similar to the one of primordial black holes below the evaporation limit  \footnote{Roughly, for  $h(r=\mathcal{R})\approx 1.1v$ we have $m<10^{11}$ kg.}. However, as they also interact through their Higgs external field, the phenomenology is expected to be distinct from the one of black holes. 
Finally, we remark that there exists an intriguing possibility that the formation of these monopoles is related to the semiclassical instability found in \cite{LMV} and discussed in terms of spontaneous scalarization in \cite{Pani}, \footnote{Note, however, that the stability analysis presented in \cite{Pani} cannot be applied to our model because the GR solution does not coexist with the monopole.}. Although for astrophysical bodies we do not expect that this instability plays a significant role, as the scalarization is negligible, it could be crucial for the formation of inflationary remnants.

\noindent \textit{Acknowledgments:} All computations were performed at the ``plate-forme technologique en calcul intensif'' (PTCI) of the University of Namur, Belgium, with financial support of the F.R.S.-FNRS (convention No. 2.4617.07. and 2.5020.11). M.R. is supported by ARC convention No.11/15-040 and S.S.  is FRIA Research Fellow.

\end{document}